\documentclass[a4paper,11pt]{article}
\usepackage{pos}
\usepackage{hyperref}
\usepackage{lineno}  

\title{Exploring the hadronic phase with momentum and azimuthal distribution of short-lived resonances and understanding the internal structure of exotic resonances with ALICE}

\author*{Hirak Kumar Koley}
\affiliation{On behalf of the ALICE collaboration}

\affiliation{Nuclear and Particle Physics Research Centre,\\
   Department of Physics, Jadavpur University, Kolkata - 700032, India}

\emailAdd{hirak.koley@cern.ch}

\abstract{
Hadronic resonances are crucial probes to understand the various phases of matter created during relativistic heavy-ion collisions. Due to their short lifetimes, the yields of these resonances can be affected by competing rescattering and regeneration mechanisms in the final hadronic phase. Rescattering can alter the momentum of the resonance decay products, limiting their reconstruction through the invariant-mass technique, while pseudo-elastic scattering can regenerate them. Final state observables such as elliptic flow, transverse momentum spectra, and measured yields of resonances could be significantly modified due to the interaction in the hadronic phase. By comparing the yields of longer-lived resonances, such as the $\phi$-meson with shorter-lived ones, such as the K$^*$(892), it is possible to obtain information about the properties and timescales of the hadronic phase. This contribution presents new Run 3 results on production yields, spectra, and flow harmonics for K$^*$(892) and $\phi$(1020) in Pb–Pb collisions at $\sqrt{s_{NN}}$ = 5.36 TeV obtained by the ALICE Collaboration. The results are compared with state-of-the-art models to interpret the underlying mechanism that can describe the experimental observations. To establish a baseline for even shorter-lived states, the $\rho(770)^{0}$ resonance ($\tau \approx 1.3$ fm$/c$) is also studied in pp collisions at $\sqrt{s}=13.6$ TeV via the $\pi^{+}\pi^{-}$ invariant-mass distribution. In addition to probe hadronic phase, the study of resonances also offers valuable insights into the non-perturbative regime of Quantum Chromodynamics (QCD). Resonances such as the f$_0$(980) and f$_1$(1285) challenge the traditional quark model. Their structure is yet unknown as they could potentially be tetraquark states or meson-meson molecules. Proposed Glueball candidates like the f$_2$(1270), f$^{'}_2$(1525), and f$_0$(1710) also provide opportunities to explore the gluonic bound states predicted by the lattice QCD. Utilizing its excellent particle identification capabilities, ALICE has recently conducted detailed studies of the exotic resonance production in pp collisions at $\sqrt{s}$ = 13 and 13.6 TeV. This contribution presents new measurements of exotic resonances such as f$_0$(980), f$_1$(1285), and the glueball candidates to get more insight into their internal structure.
}

\FullConference{The European Physical Society Conference on High Energy Physics (EPS-HEP2025)\\
7-11 July 2025\\
Marseille, France\\}

\begin{document}

\maketitle

\section{Introduction}

Ultra-relativistic heavy-ion collisions at the CERN Large Hadron Collider (LHC) create conditions of extreme temperature and energy density where a deconfined state of quarks and gluons, known as the Quark–Gluon Plasma (QGP), is formed \cite{SHURYAK198071}. These collisions provide a unique opportunity to investigate the properties of this novel state of matter predicted by Quantum Chromodynamics (QCD). Hadronic resonances play a central role in this investigation, as their short lifetimes make them highly sensitive to the conditions of the late hadronic phase. Their reconstructed yields and kinematic distributions are affected by a competition between two mechanisms: the rescattering of the decay daughters in the hadronic medium, which suppresses the final yield, and resonance regeneration through pseudo-elastic scattering, which can enhance it \cite{Bleicher2002}. Since rescattering also modifies the momentum distributions of the decay daughters, resonance reconstruction relies on the invariant-mass technique to robustly identify the parent particle.

Final-state observables such as transverse momentum ($p_{\mathrm{T}}$) spectra, particle yields, and anisotropic flow coefficients (e.g. elliptic flow $v_{2}$) provide essential constraints on the properties and timescales of the hadronic phase \cite{Acharya2018}. In particular, comparing short-lived states with longer-lived resonances, such as the $K^{*}(892)^{0}$ and $\phi$(1020), enhances sensitivity to in-medium effects. In this contribution, new Run 3 ALICE results on the production yields, $p_{\mathrm{T}}$ spectra, and flow harmonics of the $K^{*}(892)^{0}$ and $\phi$(1020) in Pb–Pb collisions at $\sqrt{s_{\mathrm{NN}}}$ = 5.36 TeV are presented and compared with state-of-the-art hydrodynamic and hadronic transport model calculations, providing insights into particle re-scattering, regeneration, and the kinetic freeze-out stage.

In addition to heavy-ion collisions, resonance studies in smaller systems probe QCD in the non-perturbative regime and test hadronization mechanisms. As part of the new results, the production of the $\rho(770)^{0}$ resonance has been measured in pp collisions at $\sqrt{s}$ = 13.6 TeV with Run 3 data.

Resonances also provide an avenue to explore exotic QCD phenomena. States with unconventional internal quark structures challenge the traditional quark model and open opportunities to study exotic candidates such as tetraquarks, meson–meson molecular states, and gluonic excitations (glueballs). Notable examples include the light scalar and axial-vector resonances $f_{0}(980)$ and $f_{1}(1285)$, whose internal composition remains debated \cite{f0}, as well as proposed glueball candidates like $f_{2}(1270)$, $f_{2}^{\prime}(1525)$, and $f_{0}(1710)$, whose masses and decay modes are constrained by lattice QCD predictions \cite{glueball}. ALICE, with its excellent tracking and particle identification, has carried out detailed measurements of these states in different collision systems. In particular, Run 2 results from pp collisions at $\sqrt{s}$ = 13 TeV, along with the new Run 3 pp results at $\sqrt{s}$ = 13.6 TeV, are presented. These studies provide new constraints on the internal structure of exotic resonances and offer valuable input for understanding the strong interaction in the non-perturbative regime.

\section{Results and Discussions}
\subsection{Production and Flow of Hadronic Resonances}

The short lifetime of hadronic resonances make them sensitive to the dynamics of the late hadronic phase. In this section, measurements of the production and azimuthal anisotropy of the $K^{*}(892)^{0}$ and $\phi$(1020) mesons in Pb–Pb collisions at $\sqrt{s_{\mathrm{NN}}}$ = 5.36 TeV are presented.

\begin{figure}
    \centering
    \includegraphics[width=0.35\linewidth, height=0.31\linewidth]{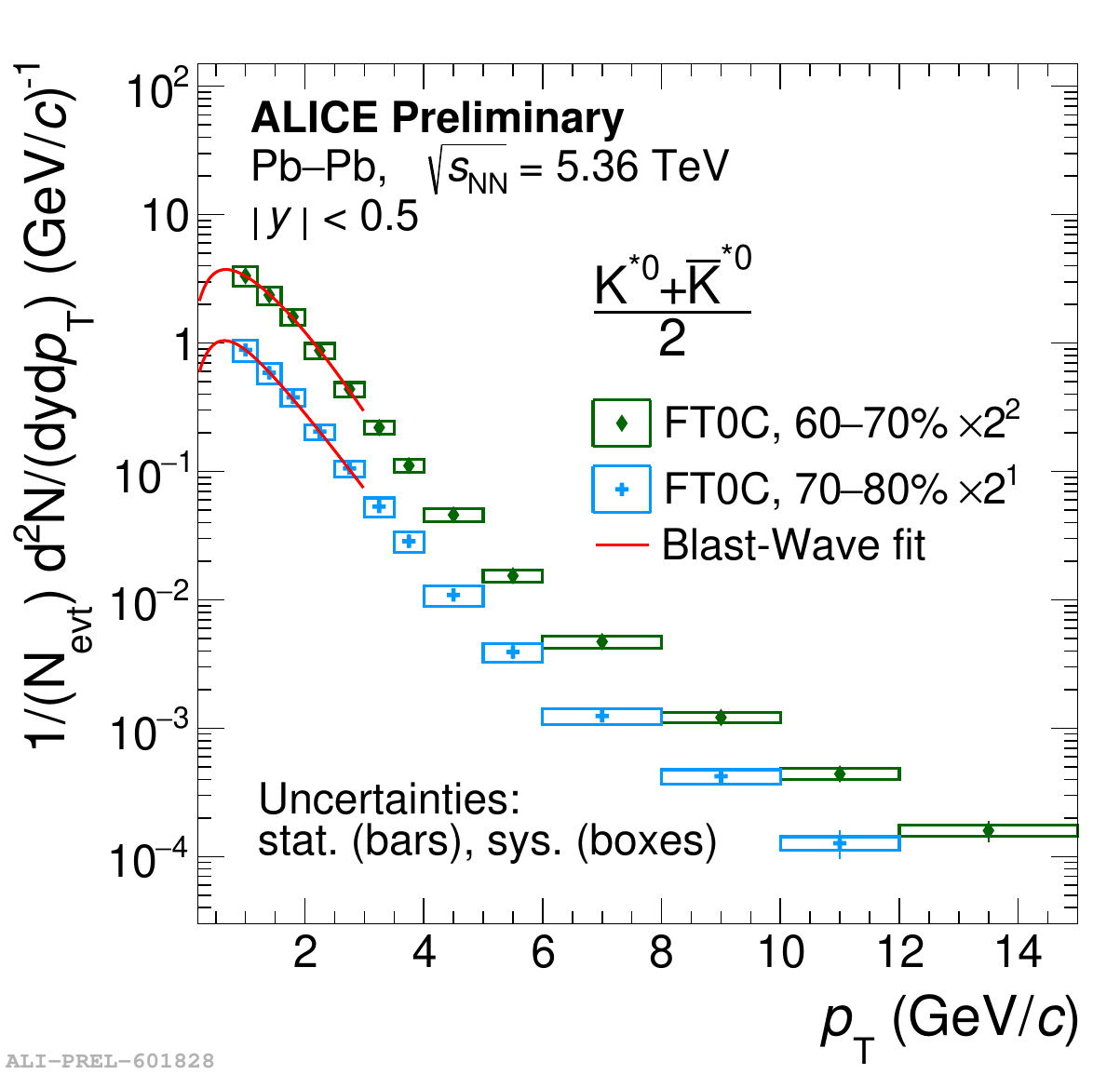}
    \includegraphics[width=0.35\linewidth, height=0.3\linewidth]{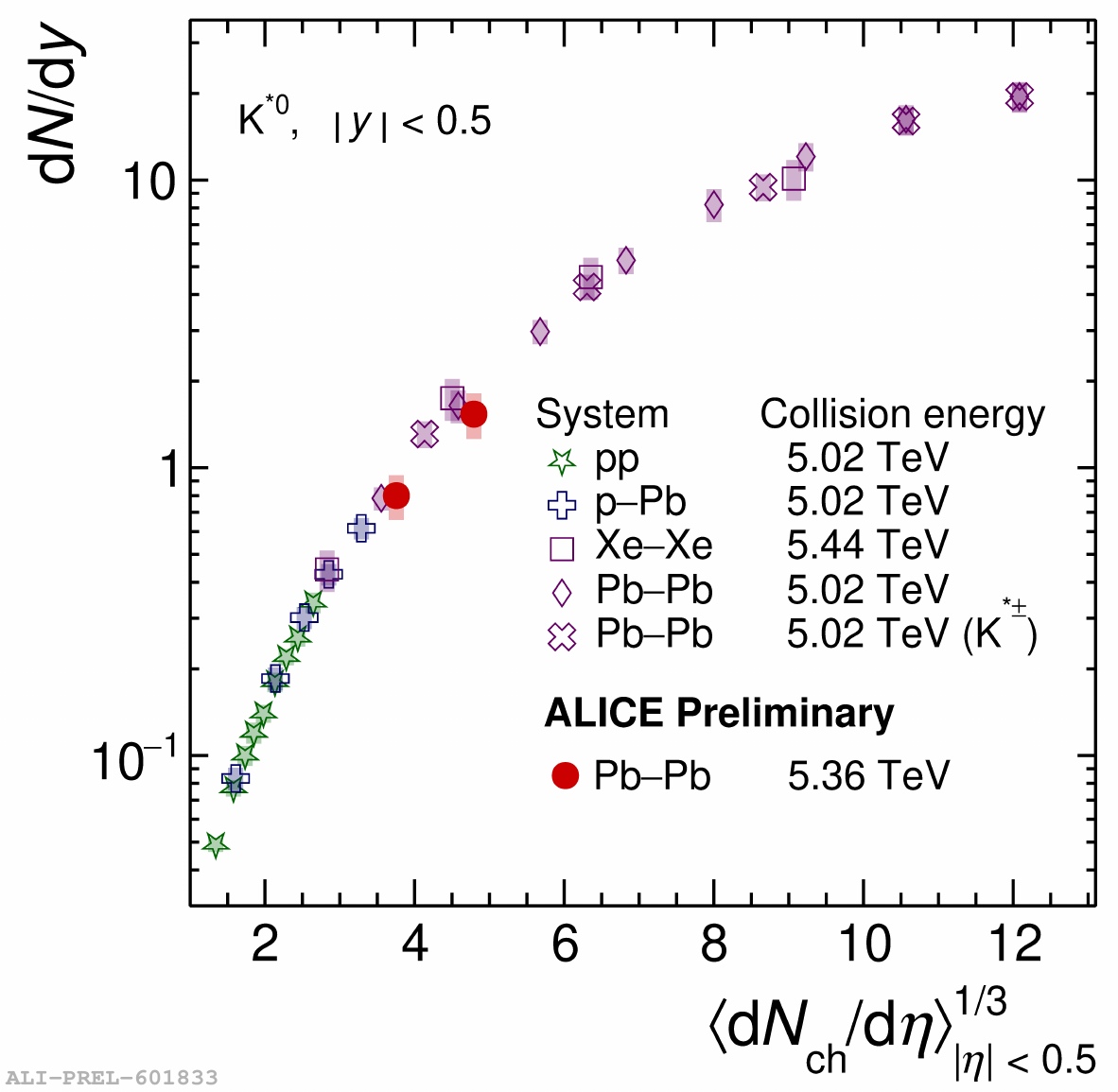}
    \caption{(left) The fitted $p_{\mathrm{T}}$ spectrum with Blast-wave function and (right) $p_{\mathrm{T}}$-integrated yield of $K^{*}(892)^{0}$ as a function of $\langle dN_{\mathrm{ch}}/d\eta \rangle^{1/3}$ in Pb-Pb collisions at $\sqrt{s_{\mathrm{NN}}}$ = 5.36 TeV.}
    \label{fig:kstar}
\end{figure}

The production of the $K^{*}(892)^{0}$ resonance is studied via its hadronic decay channel $K^{*}(892)^{0} \rightarrow K^{\pm}\pi^{\mp}$. Figure \ref{fig:kstar} (left) shows the measured $p_{\mathrm{T}}$ spectra of the $K^{*}(892)^{0}$ in different centrality classes in peripheral region. 
Fits with the Blast-Wave model reproduces the data. 
Figure \ref{fig:kstar} (right) shows the centrality dependence of the integrated yield of $K^{*}(892)^{0}$, which increases with $\langle dN_{\mathrm{ch}}/d\eta \rangle^{1/3}$ as expected from bulk particle production. 
The results are compared with Run 2 measurements, showing good consistency within uncertainties.

\begin{figure}
    \centering
    \includegraphics[width=0.4\linewidth]{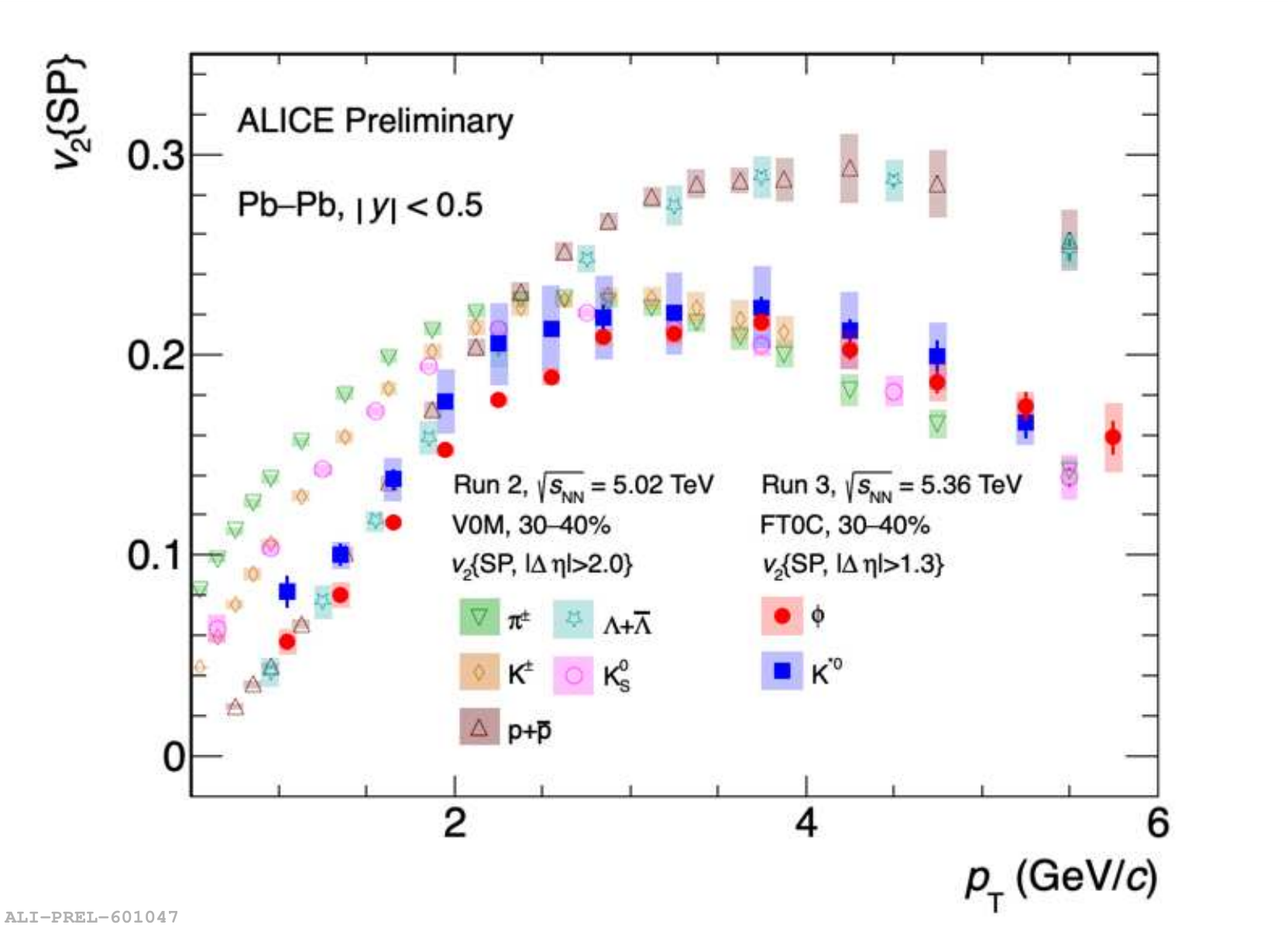}
    \includegraphics[width=0.59\linewidth, height=0.285\linewidth]{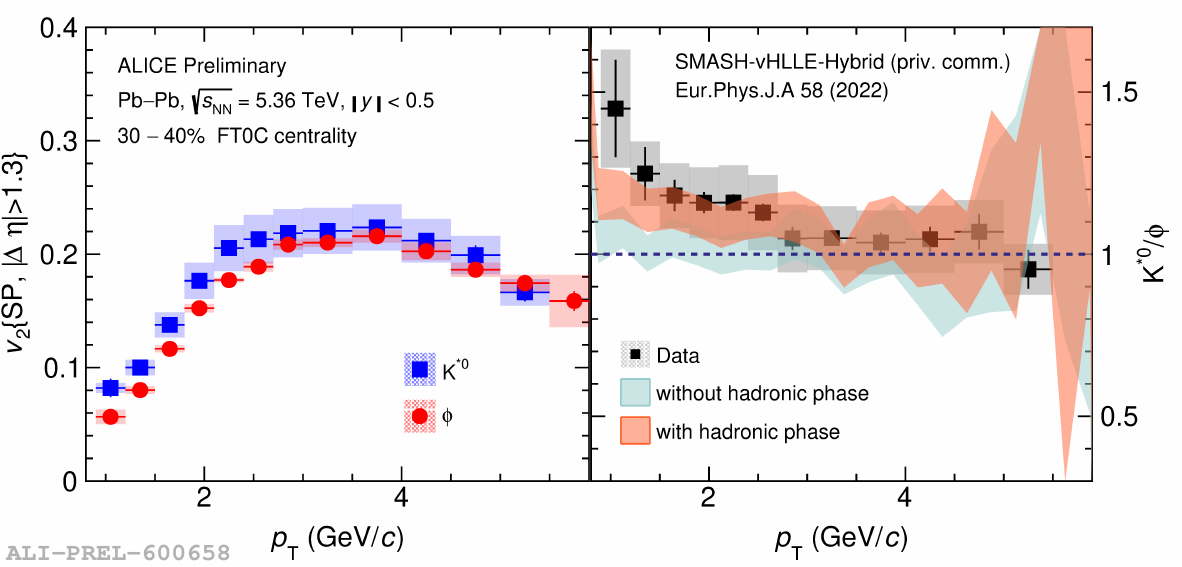}
    \caption{(left) Comparison of $v_{2}$ of $K^{*}(892)^{0}$ and $\phi$(1020) from Run 3 with that of other hadrons from Run 2 for centrality 30–40$\%$ and (right) comparison of $v_{2}$ between $K^{*}(892)^{0}$ and $\phi$(1020) with model prediction.}
    \label{fig:v2}
\end{figure}

The role of the hadronic phase is further probed through elliptic flow ($v_{2}$) measurements, determined using the Scalar Product (SP) method \cite{Acharya2018}. The comparison of $v_{2}$ between the short-lived $K^{*}(892)^{0}$ $(\tau \approx 4.16 fm/c)$ and the longer-lived $\phi$(1020) $(\tau \approx 46 fm/c)$ provides sensitivity to the duration of the hadronic stage. Figure \ref{fig:v2} (left) shows $v_{2}(p_{\mathrm{T}})$ for both resonances in the 30–40$\%$ centrality class, where they follow the expected baryon-meson grouping at intermediate $p_{\mathrm{T}}$. 
At low $p_{\mathrm{T}}$ region, the $K^{*}(892)^{0}$ exhibits systematically higher $v_{2}$ compared to the  $\phi$(1020) as shown in Fig. \ref{fig:v2} (right). 
These observations demonstrate that both the $K^{*}(892)^{0}$ and $\phi(1020)$ mesons participate in the collective expansion of the medium and hadronize predominantly via quark coalescence in the QGP, while the systematically lower $v_{2}$ of $\phi(1020)$ reflects its earlier decoupling from the hadronic medium due to its larger mass and longer lifetime, which reduces its sensitivity to rescattering.
Comparisons with the SMASH–VHLLE hybrid model demonstrate that the inclusion of a hadronic afterburner is essential to reproduce the observed differences in $v_{2}$ \cite{smash}.

To establish a baseline for even shorter-lived states, the $\rho(770)^{0}$ resonance ($\tau \approx 1.3$ fm/$c$) is studied in pp collisions at $\sqrt{s}$ = 13.6 TeV. The analysis is performed via the $\pi^{+}\pi^{-}$ invariant-mass spectrum. Figure \ref{fig:cocktail} (left) shows the distribution in the $p_{\mathrm{T}}$ interval 6.0 < $p_{\mathrm{T}}$ < 8.0 GeV/$c$. A successful cocktail fit, including contributions from the $\rho(770)^{0}$ and other background sources, demonstrates the feasibility of reconstructing such ultra-short-lived states in small systems. This measurement provides a valuable reference for future heavy-ion studies.

\subsection{Structures of Exotic Resonances and Glueball Search}

Beyond their role as probes of the hadronic phase, resonances also provide access to the non-perturbative regime of QCD, where states with unconventional quark configurations and gluonic excitations may exist \cite{scalarmeson, Jaffe}. ALICE exploits small collision systems to study the production of exotic hadron candidates and to search for possible glueball states, such as the scalar resonance $f_{0}(1710)$ \cite{glueball}.

\begin{figure}
    \centering
    \includegraphics[width=0.45\linewidth]{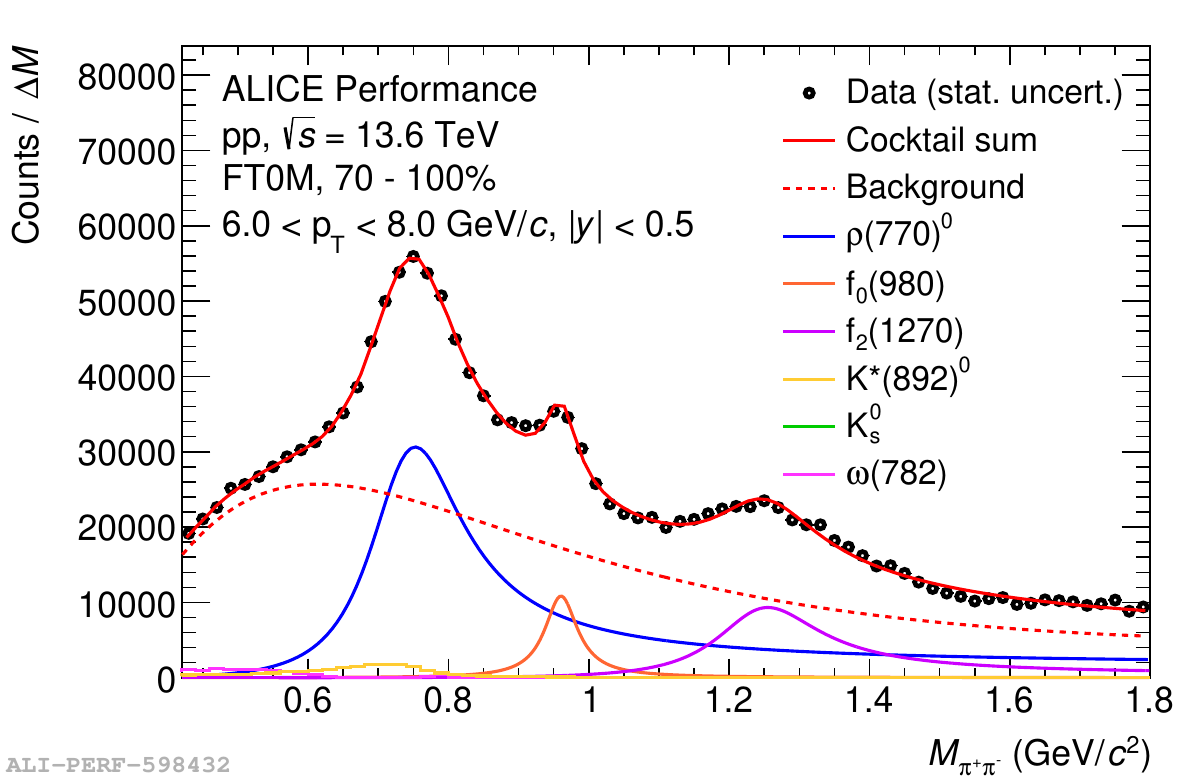}
    \includegraphics[width=0.3\linewidth, height=0.29\linewidth]{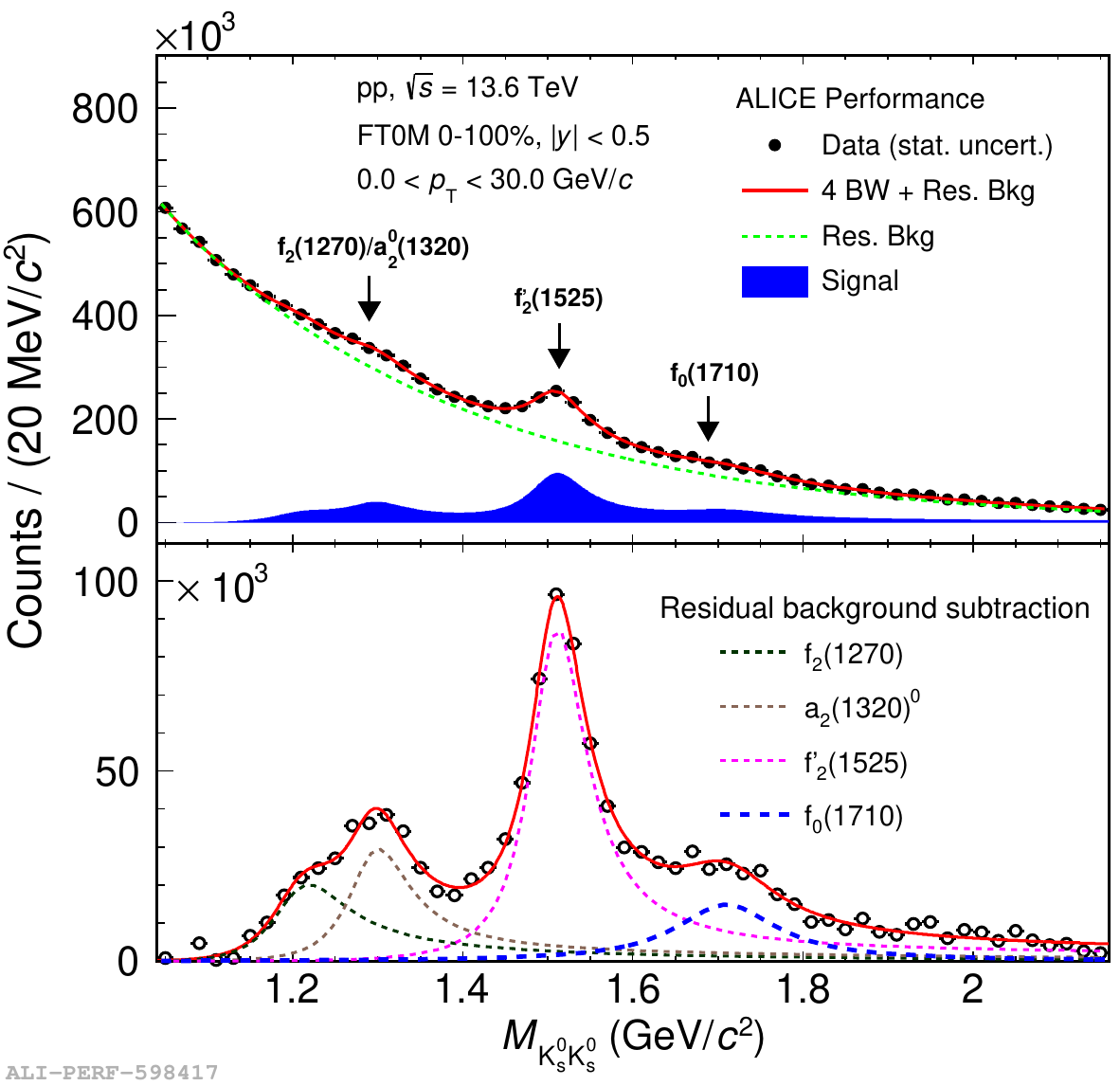}
    \caption{Invariant mass distribution of $\pi^{+}\pi^{-}$ (left) and $K_{S}^{0}K_{S}^{0}$ (right) pairs modelled using relativistic Breit–Wigner functions for resonance peaks and a smoothly varying function for residual background.}
    \label{fig:cocktail}
\end{figure}

Further insight into exotic QCD structures is obtained by investigating glueball candidates. The decay channel $f_{X} \rightarrow K_{S}^{0}K_{S}^{0}$ is used to reconstruct the tensor meson $f_{2}$(1270), its strange partner $f_{2}^{\prime}$(1525), and the scalar $f_{0}$(1710), considered a leading glueball candidate \cite{Ks0channel}. Figure \ref{fig:cocktail} (right) shows the invariant-mass distribution of $K_{S}^{0}K_{S}^{0}$ pairs in pp collisions at $\sqrt{s}$ = 13.6 TeV. Distinct peaks corresponding to the $f_{2}(1270)/a_{2}^{0}$(1320) doublet, $f_{2}^{\prime}$(1525), and $f_{0}$(1710) are clearly visible. The spectrum is fitted with relativistic Breit–Wigner functions for the resonant signals together with a residual background component.

\begin{figure}
    \centering
    \includegraphics[width=0.4\linewidth]{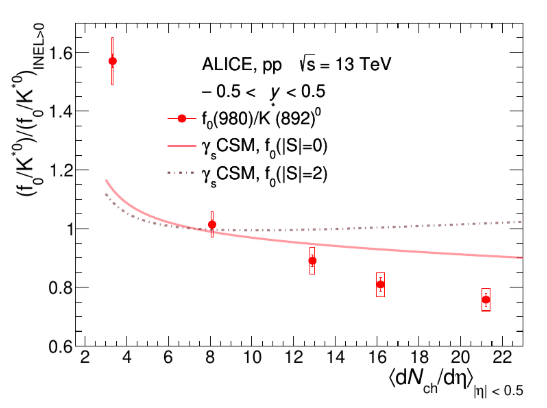}
    \includegraphics[width=0.33\linewidth, height=0.29\linewidth]{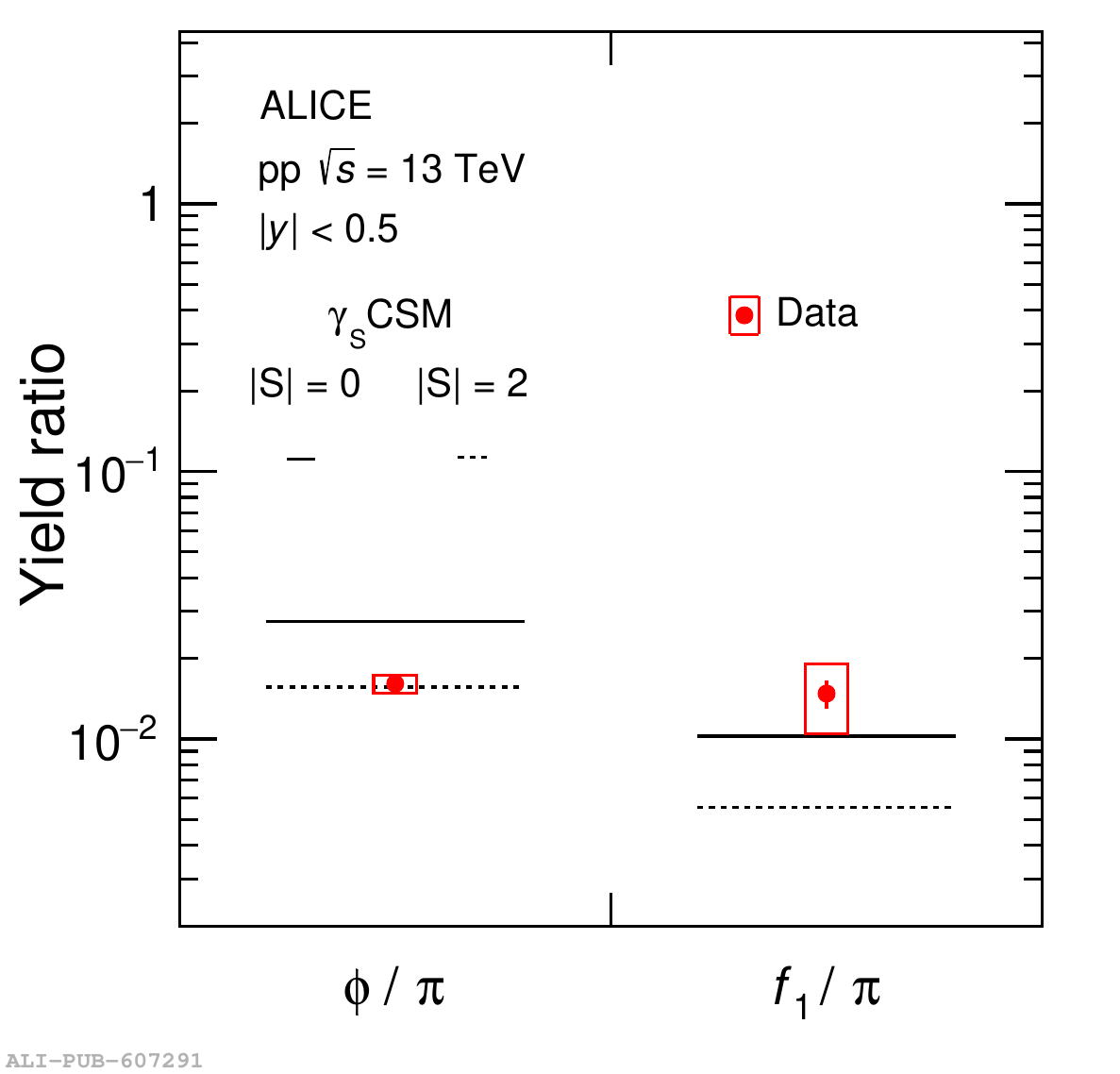}
    \caption{Comparison of $\gamma_{s}$-CSM predictions with ALICE data for (left) $f_{0}(980)/K^{*}(892)^{0}$ double ratio \cite{arxivedf0} and (right) integrated yield ratios of $\phi$ and $f_1$ with respect to $\pi$ in pp collisions at $\sqrt{s}$ = 13 TeV.}
    \label{fig:exotic}
\end{figure}

Figure \ref{fig:exotic} (left) presents the yield ratio $f_{0}(980)/K^{*}(892)^{0}$ as a function of the charged-particle multiplicity $(\langle dN_{\mathrm{ch}}/d\eta \rangle^{1/3})$ in pp collisions at $\sqrt{s}$ = 13 TeV. The $f_{0}(980)$ is of particular interest due to its debated internal structure, with possible interpretations including a conventional $q\bar{q}$ meson, a tetraquark $(qq\bar{q}\bar{q})$ state, or a $K\bar{K}$ molecular bound state \cite{f0}. The decreasing trend with multiplicity is compared to predictions from the $\gamma_{s}$-Canonical Statistical Model ($\gamma_{s}$-CSM) \cite{CSM}, where scenarios with strangeness content $|S|=0$ and $|S|=2$ are considered. The data qualitatively favor the $|S|=2$ scenario, hinting at an effective strange-quark content in the $f_{0}(980)$.

ALICE has recently identified the \(f_{1}(1285)\) in pp collisions at \(\sqrt{s} = 13\) TeV, focusing on the \(K^{0}_{\mathrm{s}}K\pi\) decay channel.  
The extracted resonance yields provide essential input for comparison with QCD-motivated models of hadron production. Yield ratios of these states to light hadrons (e.g. $\pi$) are particularly sensitive to the presence of exotic production mechanisms. Figure \ref{fig:exotic} (right) shows published ALICE results for the $\phi/\pi$ and $f_{1}/\pi$ ratios compared with $\gamma_{s}$-CSM predictions, illustrating ongoing efforts to constrain non-perturbative dynamics.
Notably, thermal model predictions for particle ratios, considering the \(f_{1}(1285)\) with strange quark content (\(|S| = 0\)), are found to be in closer agreement with experimental measurements compared to predictions assuming \(|S| = 2\). 

\section{Summary}

This contribution presented the latest ALICE results on the production and collectivity of hadronic resonances, together with new studies of exotic structures in pp collisions.
In Pb–Pb collisions at $\sqrt{s_{\mathrm{NN}}}$ = 5.36 TeV, the vector mesons $K^{*}(892)^{0}$ and $\phi(1020)$ provide sensitive probes of the hot, dense medium. In contrast, the elliptic flow $(v_{2})$ of the $\phi(1020)$ is found to be systematically lower than that of the $K^{*}(892)^{0}$ at low $p_{\mathrm{T}}$, reflecting its larger mass and longer lifetime. These observations highlight the different sensitivities of short- and long-lived resonances to the hadronic stage and demonstrate the importance of including a realistic hadronic afterburner in transport models to reproduce the measured resonance yields and flow patterns.

In small collision systems, resonance studies provide insight into the internal structure of hadrons and non-perturbative QCD phenomena. The decreasing $f_{0}(980)/K^{*}(892)^{0}$ ratio with multiplicity suggests differences in production mechanisms or sensitivity to the system size, challenging simple $\gamma_{s}$-CSM Model expectations. Invariant-mass analyses of the $K^{0}_{S}K^{0}_{S}$ decay channel yield clear signals of the $f_{2}$(1270), $f_{2}^{\prime}$(1525), and the glueball candidate $f_{0}$(1710). These measurements represent a critical step toward constraining the internal structure of exotic resonances and assessing the glueball hypothesis.
With its excellent tracking and particle-identification capabilities, ALICE continues to provide precise measurements across both heavy-ion and small collision systems, advancing our understanding of QCD matter and the complex structure of its hadronic states.

\begin{acknowledgments}
The author, Hirak Kumar Koley, gratefully acknowledges financial support from the Department of Science and Technology, Government of India, under the “Mega Facilities in Basic Science Research” scheme.
\end{acknowledgments}


\end{document}